\documentclass[10pt]{ieeeconf}

\newtheorem{thm}{Theorem}[section]
\newtheorem{cor}[thm]{Corollary}
\newtheorem{lem}[thm]{Lemma}

\newtheorem{problem}{Problem}

\newtheorem{example}{Example}

\usepackage{graphicx} 
\graphicspath{{figures/}}

\usepackage{amsmath}
\usepackage{amssymb}

\usepackage[space]{cite}
\usepackage{algorithm}
\usepackage{algorithmic}
\usepackage{multirow}
\usepackage{rotating}
\usepackage{subfigure}
\usepackage{color}

\begin{document}

\title{\LARGE \bf Distributed Control of the Laplacian Spectral Moments of a
Network}

\author{Victor~M.~Preciado,~Michael~M.~Zavlanos,~Ali~Jadbabaie,~and~George~J.~Pappas
\thanks{This work is partially supported by the ONR MURI HUNT and AFOR MURI Complex Networks programs}
\thanks{Victor~M.~Preciado,~Michael~M.~Zavlanos,~Ali~Jadbabaie,~and~George~J.~Pappas
        are with GRASP Laboratory, School of Engineering and Applied Science,
        University of Pennsylvania, Philadelphia, PA 19104, USA
        {\tt\footnotesize \{preciado,zavlanos,jadbabai,pappasg\}@seas.upenn.edu}}}


\maketitle

\begin{abstract}

It is well-known that the eigenvalue spectrum of the Laplacian
matrix of a network contains valuable information about the network
structure and the behavior of many dynamical processes run on it. In
this paper, we propose a \emph{fully decentralized} algorithm that
iteratively modifies the structure of a network of agents in order
to control the moments of the Laplacian eigenvalue spectrum.
Although the individual agents have knowledge of their \emph{local}
network structure only (i.e., myopic information), they are
collectively able to aggregate this local information and decide on
what links are most beneficial to be added or removed at each time
step. Our approach relies on gossip algorithms to distributively
compute the spectral moments of the Laplacian matrix, as well as
ensure network connectivity in the presence of link deletions. We
illustrate our approach in nontrivial computer simulations and show
that a good final approximation of the spectral moments of the
target Laplacian matrix is achieved for many cases of interest.

\end{abstract}

\section{Introduction}\label{sec:introduction}

A wide variety of distributed systems composed by autonomous agents
are able to display a remarkable level of self-organization despite the
absence of a centralized coordinator \cite{W62,H83}. For example,
the structure of many ``self-engineered'' networks, such as
social and economic networks, emerges from local interactions
between agents aiming to optimize their local utilities \cite{J08}.
Motivated by the implications of a network's Laplacian spectrum on its
structure (i.e., number of connected components) and behavior of
dynamical processes implemented on it (i.e., speed of convergence of
distributed consensus algorithms), we propose a distributed model of graph evolution in
which autonomous agents can modify their local neighborhood in order
to control a set of moments of the network Laplacian spectrum.

The eigenvalue spectra of a network provide valuable information
regarding the behavior of many dynamical processes running within
the network \cite{P08}. For example, the eigenvalue spectrum of the
Laplacian matrix of a graph affects the mixing speed of Markov
chains \cite{Ald82}, the stability of synchronization of a network
of nonlinear oscillators \cite{PC98,PV05}, the spreading of a virus
in a network \cite{DGM08,PJ09}, as well as the dynamical behavior of
many decentralized network algorithms \cite{L97}. Similarly, the
second smallest eigenvalue of the Laplacian matrix (also called
spectral gap) is broadly considered a critical parameter that
influences the stability and robustness properties of dynamical
systems that are implemented over information networks
\cite{OM04,FM04}. Optimization of the spectral gap has been studied
by several authors both in a centralized \cite{GMS90,GB06,KM06} and
decentralized context \cite{DJ06}. In contrast, our approach focuses
on controlling the moments of the Laplacian eigenvalue spectrum. In
this way, we can influence the behavior of certain dynamical
processes run within the network. As we show, the benefit of
controlling the spectral moments, and especially the lower order
ones, lies in lower computational cost and elegant distributed
implementation.

A major challenge in our approach is to efficiently control the
spectral moments of a network in a {\em fully distributed fashion}
while maintaining {\em network connectivity} in the presence of link
deletions. Our work is related to \cite{KM08}, where a fully
distributed algorithm is proposed to compute the full set of
eigenvalues and eigenvectors of a matrix representing the network
topology. However, our approach is computationally cheaper since
computation of the spectral moments does not require a complete
eigenvalue decomposition, but can be performed distributively by
averaging local network information, such as node degrees. On the other hand, control of the network structure
to the desired set of spectral moments is based on greedy actions
(link additions and deletions) that are the result of distributed
agreement protocols between the agents. We show that our distributed
topology control algorithm is stable and converges to a network with
spectral moments ``close'' to the desired. The performance of our
algorithm is illustrated in computer simulations.

The rest of this paper is organized as follows. In
Section~\ref{sec:problem}, we formulate the problem under
consideration and review some terminology. In
Section~\ref{sec:compute_moments}, we derive closed-form expressions
for the first four moments of the Laplacian spectrum in terms of
graph properties that can be measured locally. Based on these
expressions, we introduce a distributed algorithm to compute these
moments. In Section~\ref{sec:control_moments}, we propose an
efficient distributed algorithm to control of the spectral moments
of a network. Finally, in Section~\ref{sec:simulations}, we
illustrate our approach with several computer simulations.

\section{Preliminaries \& Problem Definition}\label{sec:problem}

Let $\mathcal{G}=\left( \mathcal{V},\mathcal{E}\right) $ denote a
graph on $n $ nodes, where $\mathcal{V}=\left\{ v_{1},\dots
,v_{n}\right\} $ denotes the set of nodes and $\mathcal{E}\subseteq
\mathcal{V}\times \mathcal{V}$ is the set of edges. If
$(v_{i},v_{j})\in \mathcal{E}$ whenever $(v_{j},v_{i})\in
\mathcal{E}$ we say that $\mathcal{G}$ is \emph{undirected} and call
nodes $ v_{i}$ and $v_{j}$ \emph{adjacent} (or neighbors), which we
denote by $ v_{i}\sim v_{j}$. The set of all nodes adjacent to node
$v$ constitutes the \emph{neighborhood} of node $v$, defined by
$\mathcal{N}^{v}=\{w\in \mathcal{ V}\;:\;(v,w)\in \mathcal{E}\}$,
and the number of those neighbors is called the \emph{degree} of
node $v$, denoted by $\deg v$. In this paper, we consider finite
\emph{simple} graphs, meaning that two nodes are connected by at
most one edge and self-loops are not allowed.

We define a \emph{walk} from $v_{0}$ to $v_{k}$ of length $k$ to be
an ordered sequence of nodes $\left( v_{0},v_{1},...,v_{k}\right) $
such that $ v_{i}\sim v_{i+1}$ for $i=0,1,...,k-1$. We say that a
graph $\mathcal{G}$ is \emph{connected} if there exists a walk
between every pair of nodes. If $ v_{0}=v_{k}$, then the walk is
closed. A closed walk with no repeated nodes (with the exception of
the first and last nodes) is called a \emph{cycle}. \emph{Triangles}
and \emph{quadrangles} are cycles of length three and four,
respectively. Let $d\left( v,w\right) $ denote the \emph{distance}
between two nodes $v$ and $w$, i.e., the minimum length of a walk
from $v$ to $w$. We say that $v$ and $w$ are $k$-th order neighbors
if $d\left( v,w\right) =k$ and define the $k$-th order neighborhood
of a node $v$ as the set of nodes within a distance $k$ from $v$,
i.e., $\mathcal{N}_{k}^{v}=\left\{ w\in \mathcal{V}\;:\;d\left(
v,w\right) \leq k\right\} $. A $k$-th order neighborhood, induces a
subgraph $\mathcal{G}_{k}^{v}\subseteq \mathcal{G}$ with node-set
$\mathcal{N}_{k}^{v}$ and edge-set $\mathcal{E}_{k}^{v}$ defined as
the set of edges in $\mathcal{E}$ that connect two nodes in $
\mathcal{N}_{k}^{v}$. We say that a graphical property
$P_{\mathcal{G}}$ is \emph{locally measurable within a radius} $k$
\emph{around a node} $v$ if $ P_{\mathcal{G}}$ is exclusively a
function of the neighborhood subgraph, i.e.,
$P_{\mathcal{G}}=f\left( \mathcal{G}_{k}^{v}\right) $. For example,
both the degree and the number of triangles touching a node are
locally measurable within a radius $1$. Also, the number of
quadrangles touching a node is locally measurable within a radius
$2$.

Graphs can be algebraically represented via the \emph{adjacency} and
\emph{Laplacian} matrices. The \emph{adjacency matrix} of an
undirected graph $\mathcal{G}$, denoted by
$A_{\mathcal{G}}=[a_{ij}]$, is an $n\times n$ symmetric matrix
defined entry-wise as $a_{ij}=1$ if nodes $v_{i}$ and $v_{j}$ are
adjacent and $a_{ij}=0$ otherwise.\footnote{For simple graphs with
no self-loops, $a_{ii}=0$ for all $i$.} The powers of the adjacency
matrix is related to walks in a graph. In particular we have the
following results \cite{Big93}:

\begin{lem}\label{lem:Biggs}The number of closed walks of length $\alpha $
joining node $v_{i}$ to itself is given by the $i$-th diagonal entry
of the matrix $A_{\mathcal{G}}^{\alpha }$.
\end{lem}

\begin{cor}\label{cor:Biggs}
Let $\mathcal{G}$ be a simple graph. Denote by $T_{i}$ and $Q_{i}$
the number of triangles and quadrangles touching node $v_{i}$,
respectively. Then $(A_{\mathcal{G}}) _{ii} =0$,
$(A_{\mathcal{G}}^{2})_{ii}=\deg v_{i}$, $(A_{\mathcal{G}}^{3})
_{ii}=2T_{i}$ and $\left( A_{\mathcal{G}}^{4}\right) _{ii}
=2Q_{i}+\left( \deg v_{i}\right) ^{2}+\sum_{j\in N_{i}}\left( \deg
v_{j}-1 \right)$.
\end{cor}

Arranging the node degrees on a diagonal matrix yields the degree
matrix $D_{ \mathcal{G}}=\mathrm{diag}\left( \deg v_{i}\right)$.
Then, the \emph{ Laplacian matrix} $L_{\mathcal{G}}$ of a graph
$\mathcal{G}$ can be defined by
$L_{\mathcal{G}}=D_{\mathcal{G}}-A_{\mathcal{G}}$. Let $\lambda
_{1}\leq \lambda _{2}\leq ...\leq \lambda _{n}$ be the eigenvalues
of $L_{ \mathcal{G}}$, where $\mathbf{1}$ is the vector of all ones. One can prove that $L_{\mathcal{G}}$ is positive semidefinite and $\lambda_{1}=0$. Furthermore, $\mathcal{G }$ is connected if and only if $\lambda_2>0$, or
equivalently, if $\ker L_{ \mathcal{G}} =\mbox{span} \{\mathbf{1}\}$
\cite{Big93}. As a result, we have the following well-known result:

\begin{thm}[\cite{JLM03}]\label{thm:consensus}
Consider a fixed undirected graph $\mathcal{G}$ on $n$ nodes and let
$\theta_{i}(t)\in\mathbb{R}$ denote the state variable of node $i$.
Let $\theta(t)=[\theta_i(t)]\in\mathbb{R}^n$ be the vector of all
states and assume $\dot{\theta}(t)=-L_{\mathcal{G}}\theta(t)$. Then
the network $\mathcal{G}$ is connected if and only if,
\begin{equation}\label{final_value}
\lim_{t\rightarrow \infty}
\theta(t)=\frac{1}{n}\sum_{i=1}^{n}\theta_{i}(0){\bf 1}\in
\mbox{span} \{\mathbf{1}\}.
\end{equation}
for all initial conditions $\theta(0)\in\mathbb{R}^n$.
\end{thm}

Theorem~\ref{thm:consensus} says that the graph $\mathcal{G}$ is
connected if and only if all nodes eventually reach a consensus on
their state values $ \theta_{i}(t)$, for all initial conditions.
Therefore, connectivity of a network $\mathcal{G}$ can be verified
almost surely by comparing the asymptotic state values
(\ref{final_value}) of all agents, for any random initialization.
Note that a similar result can be obtained by application of a
\emph{finite-time} maximum (or minimum) consensus \cite{Cortes08}.

\subsection{Problem Definition}

Consider a discrete-time sequence of graphs $\{\mathcal{G}
(s)\}_{s\geq 1}$ where $s\in \left\{ 1,2,\dots \right\} $ is the
discrete time index. We denote by $\{\lambda_{i}(s)\}_{s\geq 1}$ the set of Laplacian eigenvalues of $\mathcal{G}(s)$. We define the $k$-th spectral moment of the
Laplacian matrix of $ \mathcal{G}(s)$ at time $s$ as:
\begin{displaymath}
m_{k}\left( s\right) \triangleq
\frac{1}{n}\sum_{i=1}^{n}\lambda _{i}^{k}(s).
\end{displaymath}
Similarly, the $k$-th centralized spectral moment of the Laplacian can written
as:
\begin{eqnarray}\label{eqn:central_moments}
\bar{m}_{k}(s) &\triangleq &\frac{1}{n}\sum_{i=1}^{n}\left( \lambda
_{i}(s)-m_{1}\left( s\right) \right) ^{k} \nonumber \\
&=&\sum_{r=0}^{k}\binom{k}{r}\left( -1\right)
^{k-r}m_{r}(s)m_{1}^{k-r}(s).
\end{eqnarray}
The first four centralized spectral moments of the Laplacian
corresponds to the mean, variance, skewness and kurtosis of the
eigenvalue spectrum and they play a central role in this paper.
Define further the error function:
\begin{equation}
\text{CME}\left( \mathcal{G}(s)\right) =\sum_{k=0}^{4}\left[ \left(
\bar{m} _{k}\left( s\right) \right) ^{1/k}-\left( \bar{m}_{k}^{\ast
}\right) ^{1/k} \right] ^{2},  \label{eqn:CME}
\end{equation}%
where $\bar{m}_{k}^{\ast }$ denotes a given set of desired
centralized moments. Since the $k$-th moment is the $k$-th power-sum
of the Laplacian eigenvalues, we include the exponents $1/k$ in the
above error function with the purpose of assigning the same
dimension to the summands in (\ref{eqn:CME}). Then, the problem
addressed in this paper is:

\begin{problem}
\label{problem} Given an initially connected graph $\mathcal{G}(0)$,
design a distributed algorithm that iteratively adds or deletes
links from $\mathcal{G}(s)$, so that the connectivity of
$\mathcal{G}(s)$ is maintained for all time $s$ and the error
function $\text{CME}(\mathcal{G}(s))$ is locally minimized for large
enough $s$.
\end{problem}

In what follows, we first propose a distributed algorithm to
efficiently compute and update $\text{CME}(\mathcal{G}(s))$ without
any explicit eigendecomposition (Section~\ref{sec:compute_moments}).
Then, in Section~\ref{sec:control_moments}, we propose a greedy
algorithm where the most beneficial edge addition/deletion is
determined based on a distributed agreement over all possible
actions that satisfy network connectivity
(Theorem~\ref{thm:consensus}). In this framework, the time variable
$s$ increases by one whenever an action is taken (i.e., an addition
or deletion of a link). For simplicity, we assume that actions are
taken one at a time, although this assumption can be relaxed to
accommodate more complex action schemes.

\section{Distributed Computation of Spectral Moments}\label{sec:compute_moments}

In what follows, we assume that the agents in the network have very
limited knowledge of the network topology. In particular, we assume
that every agent $v$ only knows the topology of the second-order
neighborhood subgraph around it, $\mathcal{G}_{2}^{v}$. (This is the
case, for example, for many online social networks, where each
individual can retrieve a list of his friends' friends.) Then,
computing the first four Laplacian spectral moments relies on
counting the presence of certain subgraphs, such a triangles and
quadrangles, in every agent's neighborhood and averaging these
quantities via distributed average consensus. In particular, since
the matrix trace operator is conserved under diagonalization (in
general, under any similarity transformation) the first three
spectral moments of the Laplacian matrix of a graph can be written
as
\begin{eqnarray}\label{Newton Expansion}
\lefteqn{m_{k}\left( L_{\mathcal{G}}\right)
=\frac{1}{N}\,\text{tr}\,L_{\mathcal{G}}^{k}=\frac{1}{n}
\,\text{tr }\,\left( D_{\mathcal{G}}-A_{\mathcal{G}}\right) ^{k}}  \nonumber \\
&=&\frac{1}{n}\sum_{p=0}^{k}\binom{k}{p}(-1)^{p}\text{ tr~}
(A_{\mathcal{G}}^{p}D_{\mathcal{G}}^{k-p}),
\end{eqnarray}
for $k\leq 3$, where we have used the fact that the trace is preserved under cyclic
permutations (i.e., tr $ADD$=tr $DAD$= tr $DDA$). We cannot use
Newton's binomial expansion for the forth moment; nevertheless, we
may still obtain the following closed form solution:
\begin{eqnarray}\label{Expansion for M4}
\lefteqn{m_{4}\left( L_{\mathcal{G}}\right)
=\frac{1}{n}\text{tr}\left(
D_{\mathcal{G}}-A_{\mathcal{G}}\right) ^{4}}  \nonumber \\
&=&\frac{1}{n}\left[ \text{tr}\left( D_{\mathcal{G}}^{4}\right)
-4\text{tr}\left(
D_{\mathcal{G}}^{3}A_{\mathcal{G}}\right) +4\text{tr}\left( D_{\mathcal{G}}^{2}A_{\mathcal{G}}^{2}\right) \right.  \\
&&\left. +2\text{tr}\left(
D_{\mathcal{G}}A_{\mathcal{G}}D_{\mathcal{G}}A_{\mathcal{G}}\right)
-4\text{tr}\left( D_{\mathcal{G}}A_{\mathcal{G}}^{3}\right)
+\text{tr}\left( A_{\mathcal{G}}^{4}\right) \right] . \nonumber
\end{eqnarray}
Expanding the traces that appear in (\ref {Newton Expansion}) and
(\ref{Expansion for M4}) we get
\begin{displaymath}
\text{tr}\left( D_{\mathcal{G}}^{q}A_{\mathcal{G}}^{p}\right)
=\sum_{i=1}^{n}\left( \deg v_{i}\right) ^{q}\left(
A_{\mathcal{G}}^{p}\right) _{ii}
\end{displaymath}
and
\begin{eqnarray*}
\lefteqn{\text{tr}\left(
D_{\mathcal{G}}A_{\mathcal{G}}D_{\mathcal{G}}A_{\mathcal{G}}\right)
=\sum_{i=1}^{n}\sum_{j=1}^{n}\left( \deg v_{i}a_{ij}\right) \left(
\deg v_{j}a_{ji}\right)} \\
&=&\sum_{i=1}^{n}\sum_{j=1}^{n}\left( \deg v_{i}\deg v_{j}\right)
a_{ij}  =\sum_{i=1}^{n}\deg v_{i}\sum_{j\in \mathcal{N}_{i}}\deg
v_{j},
\end{eqnarray*}
which substituted back in equations (\ref{Newton Expansion}) and
(\ref{Expansion for M4}) give the following expression for $k\leq 3$
\begin{equation}\label{newton-like}
m_{k}\left( L_{\mathcal{G}}\right)
=\frac{1}{n}\sum_{i=1}^{n}\sum_{r=0}^{k}\binom{k}{r} (-1)^{r}\left(
\deg v_{i}\right) ^{k-r}\left( A_{\mathcal{G}}^{r}\right) _{ii}.
\end{equation}
For $k=4$, we can also simplify the Laplacian spectral moment, which
now becomes
\begin{eqnarray}\label{Expansion moments k4}
\lefteqn{m_{4}\left( L_{\mathcal{G}}\right)
=\frac{1}{n}\sum_{i=1}^{n}\left[ \left( \deg v_{i}\right)
^{4}-4\left( \deg v_{i}\right) ^{3}\left( A_{\mathcal{G}}\right)
_{ii}+\right. }  \nonumber  \\
&&\left. +4\left( \deg v_{i}\right) ^{2}\left(
A_{\mathcal{G}}^{2}\right) _{ii}+2\deg
v_{i}\sum_{j\in \mathcal{N}_{i}}\deg v_{j}-\right.   \nonumber \\
&&\left. -4\left( \deg v_{i}\right) \left(
A_{\mathcal{G}}^{3}\right) _{ii}+\left( A_{\mathcal{G}}^{4}\right)
_{ii}\right] .
\end{eqnarray}

Substituting the expressions for $\left( A_{\mathcal{G}}^{r}\right)
_{ii}$ from Lemma~\ref{lem:Biggs} and Corollary~\ref{cor:Biggs} in
equations (\ref {newton-like}) and (\ref{Expansion moments k4}) we
obtain the first four spectral moments of the Laplacian matrix
$L_{\mathcal{G}}$ as a function of the second-order neighborhood
subgraphs only
\begin{subequations}
\label{Moments Main Expression}
\begin{align}
m_{1}\left( L_{\mathcal{G}}\right) & =\frac{1}{n}\sum_{i=1}^{n}\deg v_{i}, \\
m_{2}\left( L_{\mathcal{G}}\right) &
=\frac{1}{n}\sum_{i=1}^{n}\left[ \left( \deg
v_{i}\right) ^{2}+\deg v_{i}\right] , \\
m_{3}\left( L_{\mathcal{G}}\right) &
=\frac{1}{n}\sum_{i=1}^{n}\,\left[ \left( \deg
v_{i}\right) ^{3}+3\left( \deg v_{i}\right) ^{2}-2T_{i}\right] , \\
m_{4}\left( L_{\mathcal{G}}\right) &
=\frac{1}{n}\sum_{i=1}^{n}\left[ \,\left( \deg v_{i}\right)
^{4}+4\,\left( \deg v_{i}\right) ^{3}+\left( \deg v_{i}\right)
^{2}-\right.  \\
& \left. \hspace{-4em}-\deg v_{i}+\left( 2\deg v_{i}+1\right)
\sum_{j\in N_{i}}\deg v_{j}-8T_{i}\deg v_{i}+2Q_{i}\right] .
\nonumber
\end{align}
\end{subequations}

Note that the expressions for the spectral moments in equations
(\ref{Moments Main Expression}) are all \emph{averages} of locally
measurable quantities (within a 2-hop neighborhood), namely, node
degrees, triangles and quadrangles touching the node. Hence, we can
apply consensus and use the result of Theorem~\ref{thm:consensus} to
obtain the first four moments in a distributed way.

\section{Distributed Control of Spectral Moments}\label{sec:control_moments}

\subsection{Possible Local Actions}\label{sec:local_actions}

The possible actions (or control variables) we consider are local
link \emph{additions} and local link \emph{deletions}. A link
addition is \emph{local} if it connects a node with another node
within its second-order neighborhood. Since agents in the network
only know their local neighborhood, a fully distributed algorithm
must limit edge additions to be local. (One could extend the
algorithm to allow connections between nodes being further than two
hops away, but this option would require much more computation and
communication.)

Let $\mathcal{N}_{1}^{i}(s)$ and $\mathcal{N}_{2}^{i}(s)$ denote the
sets of neighbors and two-hop neighbors of node $i$ at time $s\geq
0$, respectively. Since any of the two nodes adjacent to a link can
take an action to delete that link, we need to decide which of the
two nodes has the authority to delete the link. To avoid
ambiguities, we define the set of edges that node $i$ has authority
to remove as: $\mathcal{E}_{d}^{i}(s)\triangleq \left\{ \left(
i,j\right)\in\mathcal{E}(s) \;|\;j\in \mathcal{N} _{1}^{i}(s),
\;i>j\right\} $. Similarly, to disambiguate between nodes adding a
(still non-existing) link between them, we define the set of
potential edges that node $i$ can create as:
$\mathcal{E}_{a}^{i}(s)\triangleq \left\{ \left(
i,j\right)\in\mathcal{E}(s) \;|\;j\in
\mathcal{N}_{2}^{i}(s)\backslash
\mathcal{N}_{1}^{i}(s),\;i>j\right\} $\footnote{Note that since the
indices of all nodes in the network are distinct, this definition
results in a unique assignment of links to nodes.}. In other words,
node $i$ can either \emph{add} a link $\left( i,j\right) \in
\mathcal{E}_{a}^{i}(s)$, or \emph{delete} a link $\left( i,j\right)
\in \mathcal{E}_{d}^{i}(s)$. Note that link deletions may violate
network connectivity. In this case, those link deletions should be
excluded from the set of allowable actions. In the next two sections
we address the cases of link deletions and link additions
separately.

\subsubsection{Link Deletions}\label{sec:link_deletions}

Network connectivity is typically inferred from the number of
trivial eigenvalues of the Laplacian matrix. However, such
approaches are not applicable in our framework, since we assume no
global knowledge of the network topology $\mathcal{G}(s)$, but only
knowledge of local neighborhoods. Instead, we employ
finite-time-maximum consensus which is a distributed algorithm and
converges to equal values on nodes belonging to the same connected
component of a graph (Theorem~\ref{thm:consensus}). Therefore, if
deletion of a link violates connectivity, both nodes adjacent to
that link will lie in different connected components and will have
different consensus values. In what follows, we extend this idea to
simultaneously checking connectivity for all possible edge deletions
in the graph using a single high-dimensional consensus algorithm.

Consider node $j$ that has authority to remove any of the links in
the set $\mathcal{E}_{d}^{j}(s)$. Each one of these links needs to
be checked with respect to connectivity and each connectivity
verification relies on a scalar consensus update, according to
Theorem~\ref{thm:consensus}. Therefore, checking all links in
$\mathcal{E}_{d}^{j}(s)$ requires $|\mathcal{E}_{d}^{j}(s)|$
consensus updates.\footnote{We define by $|X|$ the cardinality of
the set $X$.} We associate with every link in
$\mathcal{E}_{d}^{j}(s)$ a consensus variable, and stacking all
these variables in a vector we obtain the state vector ${\bf
x}_{jj}(s)\in\mathbb{R}^{|\mathcal{E}_{d}^{j}(s)|}$. Running a
distributed consensus over the network, requires participation of
all other nodes $i\neq j$. This is possible by defining the state
variables ${\bf x}_{ij}(s)\in\mathbb{R}^{|\mathcal{E}_{d}^{j}(s)|}$.
All vectors ${\bf x}_{ij}(s)$ are initialized randomly and are
updated by node $i$
according to the following maximum consensus:\\
{\em Case I}: If $(i,j)\not\in\mathcal{E}_{d}^{i}(s)\cup
\mathcal{E}_{d}^{j}(s)$, i.e., if nodes $i$ and $j$ are not
neighbors, then
\begin{equation}\label{eqn:deletion_1}
{\bf x}_{ij}(s) := \max_{k\in\mathcal{N}_1^i(s)}\left\{{\bf
x}_{ij}(s),{\bf x}_{kj}(s)\right\},
\end{equation}
with the maximum taken elementwise on the vectors,\\
{\em Case II}: If $(i,j)\in\mathcal{E}_{d}^{j}(s)$, i.e., if nodes
$i$ and $j$ are neighbors and node $j$ has authority over link
$(i,j)$, then
\begin{equation}\label{eqn:deletion_2}
[{\bf x}_{ij}(s)]_{(i,j)} :=
\max_{k\in\mathcal{N}_1^i(s)\backslash\{j\}}\left\{[{\bf
x}_{ij}(s)]_{(i,j)},[{\bf x}_{kj}(s)]_{(i,j)}\right\},
\end{equation}
and
\begin{equation}\label{eqn:deletion_3}
[{\bf x}_{ij}(s)]_{(l,j)} :=
\max_{k\in\mathcal{N}_1^i(s)}\left\{[{\bf x}_{ij}(s)]_{(l,j)},[{\bf
x}_{kj}(s)]_{(l,j)}\right\},
\end{equation}
for $l\neq i$, where $[{\bf x}_{kj}(s)]_{(l,j)}\in\mathbb{R}$
denotes the entry of ${\bf x}_{kj}(s)$ corresponding to the link
$(l,j)$,\\
{\em Case III}: If $(i,j)\in\mathcal{E}_{d}^{i}(s)$, i.e., if nodes
$i$ and $j$ are neighbors and node $i$ has authority over link
$(i,j)$, then
\begin{equation}\label{eqn:deletion_4}
[{\bf x}_{ii}(s)]_{(i,j)} :=
\max_{k\in\mathcal{N}_1^i(s)\backslash\{j\}}\left\{[{\bf
x}_{ii}(s)]_{(i,j)},[{\bf x}_{ki}(s)]_{(i,j)}\right\}.
\end{equation}

Once consensus (\ref{eqn:deletion_1})--(\ref{eqn:deletion_4}) has
converged, node $i$ compares the entries $[{\bf x}_{ii}(s)]_{(i,j)}$
and $[{\bf x}_{ji}(s)]_{(i,j)}$ for all
$(i,j)\in\mathcal{E}_{d}^{i}(s)$. Since, violation of connectivity
due to deletion of the link $(i,j)$ would result in nodes $i$ and
$j$ being in different connected components of the network, $[{\bf
x}_{ii}(s)]_{(i,j)}=[{\bf x}_{ji}(s)]_{(i,j)}$ implies that the
reduced network is still connected. Hence, we can define a set
\begin{equation}\label{eqn:safe_deletions}
\mathcal{E}_{\sigma}^{i}(s) \triangleq
\left\{(i,j)\in\mathcal{E}_{d}^{i}(s) | [{\bf
x}_{ii}(s)]_{(i,j)}=[{\bf x}_{ji}(s)]_{(i,j)} \right\},
\end{equation}
containing the {\em safe} links adjacent to node $i$ that if
deleted, connectivity is maintained.

\subsubsection{Connectivity Verification}\label{sec:connect_verif}

\begin{algorithm}[t]
\caption{Connectivity verification} \label{alg:connect_verif}
\begin{algorithmic}[1]
\REQUIRE ${\bf x}_{ij}\in\mathbb{R}^{|\mathcal{E}_{d}^{i}|}$ for all $j\in\mathcal{V}$ ;%
\REQUIRE $T_{ij}=[0\dots 1_{i}\dots 0]^T$, $\forall \; j\in\mathcal{V}$;%
\IF{$\exists \; j\in\mathcal{V}$ such that $\min\{T_{ij}\}=0$}%
\STATE Update ${\bf x}_{ij}$ by (\ref{eqn:deletion_1})--(\ref{eqn:deletion_4});%
\STATE $T_{ij} := \max_{k\in\mathcal{N}_{i}^1}\{T_{ij},T_{kj}\}$;%
\ELSIF{$\min\{T_{ij}\}=1$ for all $j\in\mathcal{V}$}%
\STATE Update $\mathcal{E}_{\sigma}^{i}$ by (\ref{eqn:safe_deletions});%
\ENDIF%
\end{algorithmic}
\end{algorithm}

The connectivity verification of link deletions, discussed in
Section~\ref{sec:link_deletions}, is illustrated in
Alg.~\ref{alg:connect_verif}. Convergence of the finite-time
consensus (\ref{eqn:deletion_1})--(\ref{eqn:deletion_4}) is captured
by a vector of tokens $T_{ij}(s)\in\{0,1\}^{n}$, initialized as
$T_{ij}(s):=[0\dots 1_i \dots 0]^T$ for all $j\in\mathcal{V}$ and
indicating that node $i$ has initialized the consensus variables for
link deletions for which node $j$ is responsible. In particular,
when all tokens of all nodes have been collected (line 4,
Alg.~\ref{alg:connect_verif}), then consensus has converged and the
set of safe link deletions $\mathcal{E}_{\sigma}^{i}(s)$ can be
computed (line 5, Alg.~\ref{alg:connect_verif}). Note that node $i$
does not need to know the neighbor sets of its non-neighbors in the
network, neither their size. Instead, the vectors ${\bf x}_{ij}(s)$
are initialized both in values and dimension as soon as vectors
${\bf x}_{kj}(s)$ are received from a neighbor
$k\in\mathcal{N}_1^i(s)$. Clearly, ${\bf x}_{jj}(s)$ are initialized
first and then propagated in the network via maximum consensus until
the information they contain reaches node $i$.

\subsection{Most Beneficial Local Action}\label{sec:best_local_action}

As discussed in Problem~\ref{problem}, the objective of this work is
to minimize the error function $\text{CME}(\mathcal{G}(s))$. For
this we propose a greedy algorithm, which for every time $s$ selects
the action that maximizes the quantity
$\text{CME}(\mathcal{G}(s))-\text{CME}(\mathcal{G}(s+1))$, if such
an action exists, and terminates if no such action exists. By
construction, such an algorithm converges to a network that {\em
locally} minimizes $\text{CME}(\mathcal{G}(s))$, while in
Section~\ref{sec:simulations}, we show that it performs well in
practice too.

In what follows we first compute the effect of a link addition or
deletion on the four first spectral moments. Although distributed
consensus could be used to compute the new moments resulting from
each possible action, as in Section~\ref{sec:compute_moments}, this
would clearly be computationally very expensive. Instead, we can
achieve this goal locally and with minor computational overhead,
based on the observation that the addition or removal of an edge
$(i,j) $ only influences the degrees of nodes $i$ and $j$, as well
as and the triangles and quadrangles touching their neighboring
nodes. Hence, agents $i$ and $j$ can communicate to compute the net
increment in the spectral moments due to the addition or deletion of
edge $(i,j) $. In particular, we get the following expressions for
the increments in the first three moments

\begin{small}
\begin{align*}
\Delta m_{1}^{\pm \left( i,j\right) } &=\pm \frac{2}{n}, \\
\Delta m_{2}^{\pm \left( i,j\right) } &=\frac{2}{n}\left[ 1\pm
\left( d_{i}+d_{j}+1\right) \right], \\
\Delta m_{3}^{\pm \left( i,j\right) } &=\frac{1}{n}\left[ \left(
3\pm 6\right) (d_{i}+d_{j})\pm 3(d_{i}^{2}+d_{j}^{2})+\left( 6\pm
2\right)\mp 6T_{ij}\right]
\end{align*}
\end{small}

\hspace{-1em}where the notation $\pm \left( i,j\right) $ indicates a
link addition $(+)$ or deletion $(-)$ and the dependence on time $s$
has been dropped for simplicity. (Similarly, one can obtain a
complicated closed-form expression for $\Delta m_{4}^{\pm \left(
i,j\right) }$, which we omit due to space limitations.) Then,
agent's $i$ copy of the $k$-th spectral moment $ m_{k}^{i}(s)$
becomes
\begin{displaymath}
m_k^{\pm(i,j)}(s)=m_k^i(s) + \Delta m_k^{\pm(i,j)}(s),
\end{displaymath}
and the associated centralized moment $\bar{m}_k^{\pm(i,j)}(s)$ can
be computed using (\ref{eqn:central_moments}). Then, for all
possible actions discussed in Section~\ref{sec:local_actions}, agent
$i$ computes the error function
\begin{equation*}
\text{CME}_{\pm(i,j)}(s) =\sum_{k=0}^{4}\left[ \left(
\bar{m}_{k}^{\pm(i,j)}(s) \right) ^{1/k}-\left( \bar{m}_{k}^{\ast
}\right) ^{1/k}\right] ^{2}.
\end{equation*}
Then, the local {\em most beneficial action} to the target
centralized moments, namely, the action that results in the maximum
decrease in the error function $\text{CME}(\cdot)$, can be defined
by\footnote{The $\max$ in the expression bellow indicates that in
case of ties in the $\min$, the highest index node wins.}
\begin{displaymath}
\nu_i(s)\triangleq \max\left\{\underset{j}{\rm argmin} \;
\left(\text{CME}_{\pm(i,j)}(s)-\text{CME}_i(s)\right)\right\},
\end{displaymath}
where $\text{CME}_i(s)$ denotes agent $i$'s copy of $\text{CME}(s)$,
and the largest decrease in the error associated with action
$\nu_i(s)$ becomes:
\begin{displaymath}
\text{CME}_{i}(s) = \text{CME}_{\pm(i,\nu_i(s))}(s)
\end{displaymath}
if $\min_{j}(\text{CME}_{\pm(i,j)}(s)-\text{CME}(s))\leq 0$ and
$\text{CME}_{i}(s) = D$, otherwise. Note that $\text{CME}_{i}(s)$ is
nontrivially defined only if the exists a link adjacent to node $i$
that if added or deleted decreases the error function
$\text{CME}(\cdot)$. Otherwise, a large value $D>0$ is assigned to
$\text{CME}_{i}(s)$ to indicate that this action is not beneficial
to the final objective. We can include all information of a best
local action in the vector
\begin{displaymath}
{\bf v}_i(s) \triangleq \left[i \; \nu_i(s) \; \text{CME}_{i}(s) \;
\bar{\bf m}_{\pm(i,\nu_i(s))}(s)\right]^T \in\mathbb{R}^{7}
\end{displaymath}
containing the local best action $(i,\nu_i(s))$, the associated
distance to the desired moments $\text{CME}_{i}(s)$, and the vector
of centralized moments $\bar{\bf m}_{\pm(i,\nu_i(s))}(s)$ due to
this action. In the following section we discuss how to compare all
local actions ${\bf v}_i(s)$ for all nodes $i\in\mathcal{V}$ to
obtain the one that decreases the distance to the desired moments
the most.

\subsection{From Local to Global Action}\label{sec:best_global_action}

To obtain the overall most beneficial action, all local actions need
to be propagated in the network and compared against each other. For
this we require minimal storage and communication as well as no
propagation of any information regarding the network topology. As in
Section~\ref{sec:local_actions}, we achieve this goal using a finite
time minimum consensus algorithm.

\begin{algorithm}[t]
\caption{Globally most beneficial action} \label{alg1}
\begin{algorithmic}[1]
\REQUIRE ${\bf v}_{i}\triangleq[i \; \nu_i \; \text{CME}_i \; \bar{\bf m}_{\pm(i,\nu_i)}]^{T}$;%
\REQUIRE $T_{i}:=[0\dots 1_{i}\dots 0]^T$;%
\IF{$\min\{T_{i}\}=0$}%
\STATE ${\bf v}_{i} := {\bf v}_{j}$, with $j=\max\{{\rm
argmin}_{k\in\mathcal{N}_1^i}\{[{\bf v}_{i}]_3,[{\bf v}_{k}]_3\}$;%
\STATE $T_{i} := \max_{j\in\mathcal{N}_1^i}\{T_{i},T_{j}\}$;%
\ELSIF{$\min\{T_{i}\}=1$ and $[{\bf v}_{i}]_3<D$}%
\STATE Update $\mathcal{N}_i$, $\bar{\bf m}_i$ and $\text{CME}_i$ according to (\ref{eqn:update_neighbors_1})--(\ref{eqn:update_CME});%
\ELSIF{$\min\{T_{i}\}=1$ and $[{\bf v}_{i}]_3=D$}%
\STATE No beneficial action. Algorithm has converged;%
\ENDIF%
\end{algorithmic}
\end{algorithm}

In particular, the desired local actions ${\bf v}_i(s)$ are
propagated in the network, along with vectors of tokens
$T_{i}(s)\in\{0,1\}^{n}$, initialized as $T_{i}(s):=[0\dots 1_i
\dots 0]^T$, indicating that node $i$ has transmitted its desired
action. During every iteration of the algorithm, every node $i$
communicates with its neighbors and updates its vector of tokens
$T_{i}(s)$ (line 3, Alg.~\ref{alg1}), as well as its desired action
${\bf v}_i(s)$ with the action ${\bf v}_j(s)$ corresponding to the
node $j$ that contains the smallest distance to the target moments
$[{\bf v}_{j}(s)]_3$, i.e.,
\begin{displaymath}
j\in{\rm argmin}_{k\in\mathcal{N}_{1}{i}(s)}\{[{\bf
v}_{i}(s)]_3,[{\bf v}_{k}(s)]_3\}.
\end{displaymath}
In case of ties in the distances to the targets $[{\bf
v}_{j}(s)]_3$, i.e., if ${\rm
argmin}_{k\in\mathcal{N}_{1}^{i}(s)}\{[{\bf v}_{i}(s)]_3,[{\bf
v}_{k}(s)]_3\}$ contains more than one nodes, then the node $j$ with
the largest label is selected (line 2, Alg.~\ref{alg1}). Note that
line 2 of Alg.~\ref{alg1} is essentially a minimum consensus update
on the entries $[{\bf v}_{i}(s)]_3$ and will converge to a common
outcome for all nodes when they have all been compared to each
other, which is captured by the condition $\min_{j=1}^{n}
T_{ij}(s)=1$ (lines 4 and 6, Alg.~\ref{alg1}). When the consensus
has converged, if there exists a node whose desired action decreases
the distance to the target moments, i.e., if $[{\bf v}_{i}(s)]_3<D$
(line 4, Alg.~\ref{alg1}), then Alg.~\ref{alg1} terminates with a
greedy action and node $i$ updates its set of neighbors
$\mathcal{N}_1^i(s)$ and vector of centralized moments $\bar{\bf
m}_i(s)$ (line 5, Alg.~\ref{alg1}). If the optimal action is a link
addition, i.e., if $[{\bf v}_{i}(s)]_2\not\in\mathcal{N}_1^i(s)$,
then
\begin{equation}\label{eqn:update_neighbors_1}
\mathcal{N}_1^i(s+1):=\mathcal{N}_1^i(s)\cup\left\{[{\bf
v}_{i}(s)]_2\right\}.
\end{equation}
On the other hand, if the optimal action is a link deletion, i.e.,
if $[{\bf v}_{i}(s)]_2\in\mathcal{N}_1^i(s)$, then
\begin{equation}\label{eqn:update_neighbors_2}
\mathcal{N}_1^i(s+1):=\mathcal{N}_1^i(s)\backslash\left\{[{\bf
v}_{i}(s)]_2\right\}.
\end{equation}
In all cases, the centralized moments and error function are updated
by
\begin{equation}\label{eqn:update_moments}
\bar{\bf m}_i(s+1) := [{\bf v}_{i}(s)]_{4:7}
\end{equation}
and
\begin{equation}\label{eqn:update_CME}
\text{CME}_i(s+1) := [{\bf v}_{i}(s)]_{3},
\end{equation}
respectively, where $[{\bf v}_{i}(s)]_{4:7} = \left[[{\bf
v}_{i}(s)]_4 \dots [{\bf v}_{i}(s)]_{7}\right]^T$. Finally, if all
local desired actions increase the distance to the target moments,
i.e., if $[{\bf v}_{i}(s)]_3=D$ (line 6, Alg.~\ref{alg1}), then no
action is taken and the algorithm terminates with a network topology
with almost the desired spectral properties. This is because no
action exists that can further decrease the distance to the target
moments.

\subsection{Synchronization}\label{sec:synchronization}

\begin{figure}[t]
  \centering
  \includegraphics[width=3in]{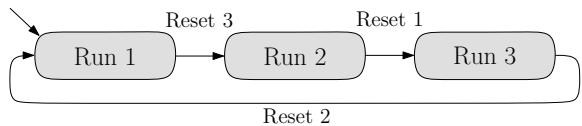}
  \caption{Synchronization: Assume node $i$ is in
  run 1. Necessary for node $i$ to transition to run 2 is
  that all other nodes are also in run 1, since
  otherwise node $i$ will be missing tokens from the nodes that are
  not in run 1 yet (currently in run 3) and Alg.~\ref{alg1}
  will not be able to converge. Once node $i$ transitions to run
  2, it initializes all variables for that run with the latest
  values from run 1, while it maintains the variables of run 1
  for nodes that are still in run 1 and it clears all variables
  of run 3 since, no node is in this run any more.}
  \label{fig:synchronization}
\end{figure}

Communication time delays, packet losses, and the asymmetric network
structure, may result in runs of the algorithm starting
asynchronously, outdated information being used for future
decisions, and consequently, nodes reaching different decisions for
the same run. In the absence of a common global clock, the desired
synchronization is ideally {\em event triggered}, where by a
triggering event we understand the time instant that a message ${\rm
Msg}[i]$ has been received by any of node $i$'s neighbors
$j\in\mathcal{N}_{1}^{i}$. We achieve such a synchronization by
labeling every algorithm run in the set $\{1,2,3\}$ and requiring
that all information exchange takes place among neighbors that are
in equally labeled runs \cite{ZP08}. Essentially, ``fast'' nodes
wait for their ``slower'' peers and, hence, all nodes are always
synchronized in the sequence $\{1,2,3,1,2,3,\dots\}$
(Fig.~\ref{fig:synchronization}).

\section{Numerical Simulations}\label{sec:simulations}

In this section we illustrate our algorithm with several numerical
examples.

\begin{figure}[tbp]
\centering
\includegraphics[width=0.9\linewidth]{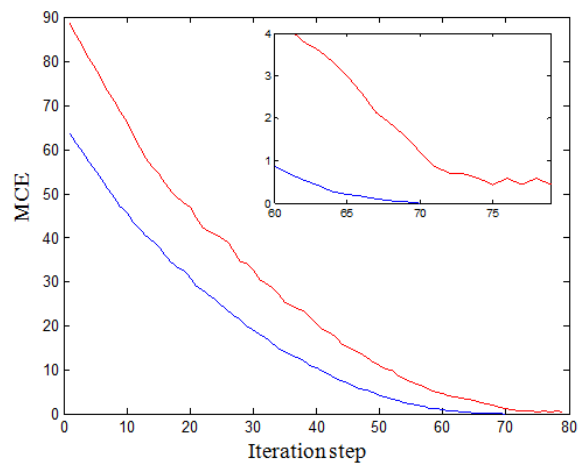}
\caption{Convergence of the error function $\text{CME}(\cdot)$ for
the star graph (blue plot) and the two-stars graph (red plot). The
subgraph in the upper right corner shows the behavior of the error
function in a neighborhood of zero. Observe that our algorithm can
match the first four moments of the star network with zero error in
finite time, but can not exactly match the moments of a the
two-stars graph, although the final error is very small.}
\label{fig:star_network}
\end{figure}

\begin{example}[Star networks]
\label{exmp:stars} Consider a star network on $10$ nodes. The first
four central moments of the associated Laplacian matrix are:
$\bar{m}_{1}^{\ast }=1.8$, $\bar{m}_{2}^{\ast }=7.56$,
$\bar{m}_{3}^{\ast }=54.14$, and $\bar{m} _{4}^{\ast }=453.49$. Our
objective is to control the topology of a randomly initialized
network on 10 nodes so that it eventually has the same set of
moments as the given star network. We observe in Fig.
\ref{fig:star_network} that our algorithm decreases the error
function (blue line) to zero in finite time. Similar performances
are observed when we repeat this procedure for star networks of any
size. Furthermore, although we are controlling the first four
moments solely, the resulting network structures are exactly the
star topologies whose moments we were trying to approximate. The
perfect reconstruction observed in this case is due to the fact that
a star graph is a extreme case in which the graph topology is
uniquely defined by their eigenvalue spectrum. Moreover, if each
agent in a star network has access to its second-order neighborhood,
it has access to the complete star topology.
\end{example}

\begin{figure}[tbp]
\centering
\includegraphics[width=0.9\linewidth]{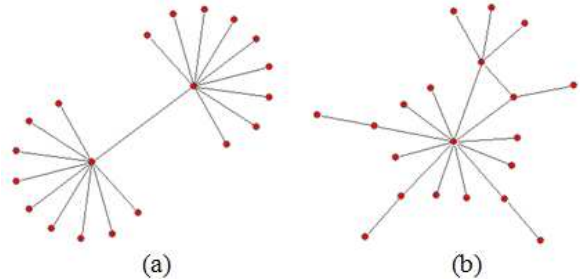}
\caption{Structures of the two-stars network (a) and the network
returned by our algorithm (b).} \label{fig:2_star_network}
\end{figure}

\begin{figure}[tbp]
\centering
\includegraphics[width=0.9\linewidth]{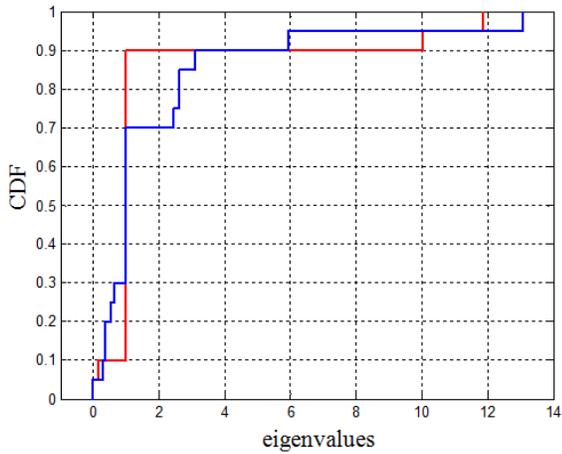}
\caption{Empirical cumulative distribution functions for the
eigenvalues of the two-stars graph (blue) and the graph returned by
our algorithm (red).} \label{fig:CDFs}
\end{figure}

\begin{example}[Two-stars network]
\label{exmp:2stars} Although our approach works very well for star
networks, the case of two-stars networks points out one of its
weaknesses, namely, its limitation in modeling network communities.
In this example, we consider two star graphs on 10 nodes each, and
connect their two central hubs with a
link. The resulting graph is the two-stars graph shown in Fig.~\ref%
{fig:2_star_network}(a). As before, we initialize our algorithm with
a random graph on 20 nodes and try to approximate the first four
central moments of the two-stars graph. In
Fig.~\ref{fig:star_network}, we observe that the error function (red
line) quickly reaches a neighborhood of zero but does not reach zero
exactly. Therefore, although our algorithm tries to generate the two
hubs in the two-star network, its local nature will not allow it
recover the highly-structured two-stars graph. Instead, it returns
the final network shown in Fig.~\ref{fig:2_star_network}(b).
Nevertheless, the eigenvalue spectra of the desired two-star
network and the network
in Fig.~\ref{fig:2_star_network}(b) are still very similar, as shown in Fig.~%
\ref{fig:CDFs}.
\end{example}

\begin{figure}[tbp]
\centering
\includegraphics[width=0.9\linewidth]{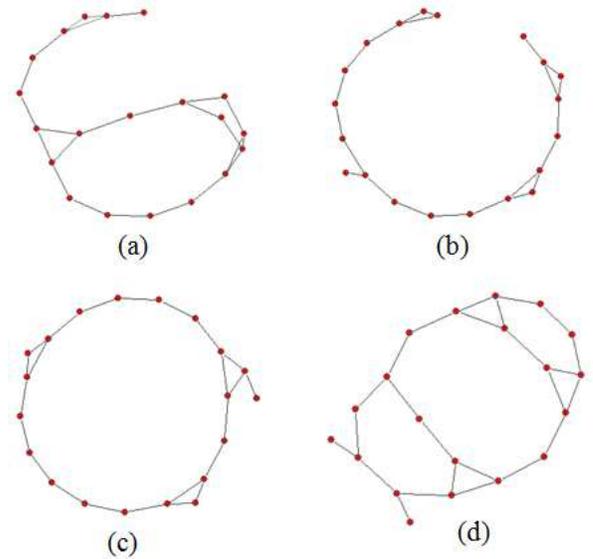}
\caption{Networks returned by our algorithm when trying to match the
first four central moments of a ring on 20 nodes.}
\label{fig:ring_networks}
\end{figure}

\begin{example}[Chain vs. ring networks]
\label{exmp:chains_rings} The objective of this example is to
illustrate how two structurally very similar (but topologically
different) target graphs, such as a chain and a ring, may affect the
performance of our algorithm. In particular, if we run our algorithm
to control the first four centralized moments of an initially random
graph towards the moments of a chain graph, we observe how the error
function converges exactly to zero in finite time. Furthermore, the
final result of our algorithm is an exact reconstruction of the
chain graph. Nevertheless, when transforming the target graph from a
chain graph into a ring graph (by adding a single link), an exact
reconstruction is very difficult. In Fig.~\ref{fig:ring_networks},
we illustrate some graphs returned by our algorithm for different
initial conditions when we control the set of moments toward the
moments of a ring network on 20 nodes. Observe that, although the
algorithm tends to create long cycles and the majority of nodes have
degree two, it fails to recreate the exact structure of the ring
graph due to the local nature of the algorithm (as in
Example~\ref{exmp:2stars}). However, although the structure of the
resulting networks is not exactly the desired ring graph, their
spectral properties are remarkably close to those of a ring. In Fig.~\ref%
{fig:CDF_ring}, we illustrate the empirical cumulative distribution
functions of the eigenvalues of the ring graph (blue plot), versus
the four empirical cumulative distribution functions corresponding
to the graphs in Fig.~\ref{fig:ring_networks}.
\end{example}

\begin{figure}[tbp]
\centering
\includegraphics[width=0.9\linewidth]{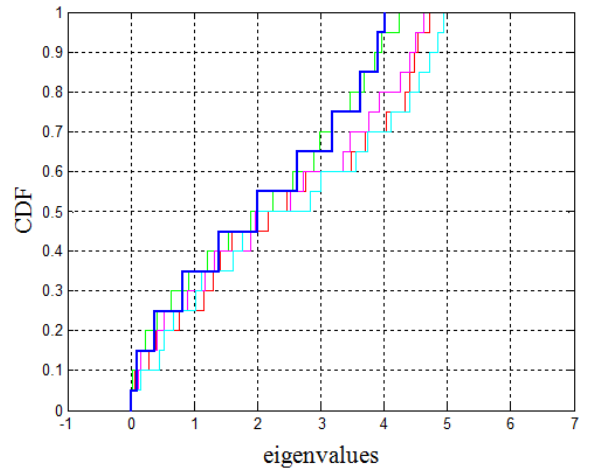}
\caption{Empirical cumulative distribution of eigenvalues for the
ring graph
with 20 nodes (blue plot) and the graphs in Fig.~\protect\ref%
{fig:ring_networks}(a) (red),
Fig.~\protect\ref{fig:ring_networks}(b) (green),
Fig.~\protect\ref{fig:ring_networks}(c) (magenta) and Fig.~\protect
\ref{fig:ring_networks}(d) (cyan).} \label{fig:CDF_ring}
\end{figure}

\begin{example}[Small-Worlds]
\label{exmp:small_world} In our final example, we use our algorithm
to control the moments of a randomly generated graph to those of a
small-world network. We consider small-world graphs as defined in
\cite{WS98}, namely, we take a ring of $n$ nodes, and connect each
node in the ring with all the nodes in its 3-hop neighborhood. Then,
we randomly rewire a fraction of the
resulting edges with probability $p$ as proposed by Watts and Strogatz \cite%
{WS98}. Our objective is to approximate the first four centralized
moments of a random instance of a small world graph with $n=40$
nodes and link probability $p=1/n$. We observed a fast convergence
of the error to a neighborhood of zero, i.e., $\text{CME}(27\text{
iterations})=.0009$, which suggests (although does not guarantee) a
good approximation between the spectra of the target small-world
graph and the graph returned by our algorithm. We repeated this
process for a link probability $p=4/n$ and similar results were
obtained. We should note, however, that although the spectral
properties between the target small-world graphs and the graphs
returned by our algorithm are remarkably similar (Fig.~\ref%
{fig:CDF_small_world}), the resulting structures are not isomorphic, as in Example~\ref{exmp:chains_rings}.
\end{example}

\begin{figure}[tbp]
\centering
\includegraphics[width=0.9\linewidth]{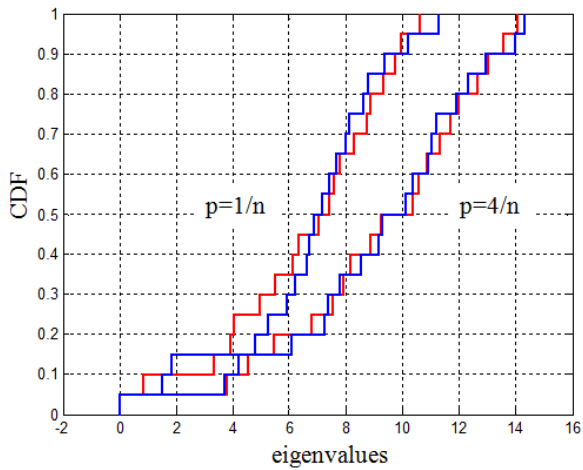}
\caption{Empirical cumulative distrubution of eigenvalues for the
small-world graphs in the example for $p=1/n$ (left) and $p=4/n$
(right).} \label{fig:CDF_small_world}
\end{figure}

\section{Conclusions and Future Research}

In this paper, we have described a fully decentralized algorithm
that iteratively modifies the structure of a network of agents with
the objective of controlling the spectral moments of the Laplacian
matrix of the network. Although we assume that each agent has access
to local information regarding the graph structure, we show that the
group is able to collectively aggregate their local information to
take a global optimal decision. This decision corresponds to the
most beneficial link addition/deletion in order to minimize an error
function that involves the first four Laplacian spectral moments of
the network. The aggregation of the local information is achieved
via gossip algorithms, which are also used to ensure network
connectivity throughout the evolution of the network.

Future work involves identifying sets of spectral moments that are
reachable by our control algorithm. (We say that a sequence of
spectral moments is reachable if there exists a graph whose moments
match the sequence of moments.) Furthermore, we observed that
fitting a set of low-order moments does not guarantee a good fit of
the complete distribution of eigenvalues. In fact, there are
important spectral parameters, such as the algebraic connectivity,
that are not captured by a small set of spectral moments.
Nevertheless, we observed in numerical simulations that fitting the
first four moments of the eigenvalue spectrum often achieves a good
reconstruction of the complete spectrum. Hence, a natural question
is to describe the set of graphs most of whose spectral information
is contained in a relatively small set of low-order moments.



\end{document}